\documentclass[aps,prl,article,twocolumn,preprintnumbers,amsmath,amssymb,superscriptaddress]{revtex4-2}

\date{\today}
\usepackage{epsfig}
\usepackage{cases}
\usepackage{subfigure}
\usepackage{amsmath}
\usepackage{graphicx}
\usepackage{dcolumn}
\usepackage{physics}
\usepackage{bm}
\usepackage[section]{placeins}
\usepackage{float}
\usepackage{mathtools}
\usepackage{dblfloatfix}
\usepackage{color}
\usepackage{comment}
\usepackage{bbm}
\usepackage[normalem]{ulem}
\usepackage[section]{placeins}
\usepackage{braket}

\usepackage{xr-hyper}

\usepackage{xr}
\usepackage[colorlinks=true, citecolor=blue, linkcolor=blue, urlcolor=blue]{hyperref}

\begin{document}
\newcommand{\su}{\uparrow}
\newcommand{\sdn}{\downarrow}
\newcommand{\HXXZ}{\hat{\mathcal{H}}_{XXZ}}
\newcommand{\Jeff}{J^z_\mathrm{eff}}
\newcommand{\psiG}{{\psi}_0}
\newcommand{\Oop}{\hat{\mathcal{O}}}
\newcommand{\FQ}{\mathcal{F}_Q}
\newcommand{\rhoDM}{\hat{\rho}}
\newcommand{\Lb}{\hat{\mathcal{L}}}
\newcommand{\Ham}{\hat{\mathcal{H}}}
\newcommand{\cop}{\hat{c}}

\title{Witnessing Nonequilibrium Entanglement Dynamics in a Strongly Correlated Fermionic Chain}

\newcommand{\affiliationHarvard}{
Department of Physics, Harvard University, Cambridge, Massachusetts 02138, USA
}

\newcommand{\affiliationMPSD}{
Max Planck Institute for the Structure and Dynamics of Matter, Center for Free-Electron Laser Science (CFEL),
Luruper Chaussee 149, 22761 Hamburg, Germany
}

\newcommand{\affiliationRWTH}{
Institut f\"ur Theorie der Statistischen Physik, RWTH Aachen University, 52056 Aachen, Germany and JARA-Fundamentals of Future Information Technology, 52056 Aachen, Germany
}

\newcommand{\affiliationPenn}{Department of Physics and Astronomy, University of Pennsylvania, Philadelphia, PA 19104, USA}

\author{Denitsa R. Baykusheva}
\affiliation{\affiliationHarvard}

\author{Mona H. Kalthoff}
\affiliation{\affiliationMPSD}

\author{Damian Hofmann}
\affiliation{\affiliationMPSD}

\author{Martin Claassen}
\affiliation{\affiliationPenn}

\author{Dante M. Kennes}
\affiliation{\affiliationRWTH}
\affiliation{\affiliationMPSD}

\author{Michael A. Sentef}
\affiliation{\affiliationMPSD}

\author{Matteo Mitrano}
\affiliation{\affiliationHarvard}

\date{\today}

\begin{abstract}
Many-body entanglement in condensed matter systems can be diagnosed from equilibrium response functions through the use of entanglement witnesses and operator-specific quantum bounds. Here, we investigate the applicability of this approach for detecting entangled states in quantum systems driven out of equilibrium. We use a multipartite entanglement witness, the quantum Fisher information, to study the dynamics of a paradigmatic fermion chain undergoing a time-dependent change of the Coulomb interaction. Our results show that the quantum Fisher information is able to witness distinct signatures of multipartite entanglement both near and far from equilibrium that are robust against decoherence. We discuss implications of these findings for probing entanglement in light-driven quantum materials with time-resolved optical and x-ray scattering methods.
\end{abstract}

\maketitle

\textit{Introduction.}--- A defining feature of quantum mechanics is the existence of nonlocal correlations \textcolor{black}{and entanglement between} distinct physical objects \cite{Bell1987speakable,Horodecki2009quantum}. Entanglement is ubiquitous in the study of quantum many-body phenomena, encompassing areas such as quantum gravity \cite{Nishioka2009holographic}, quantum information \cite{Bouwmeester2000the,nielsen2010quantum}, and condensed matter physics \cite{Amico2008entanglement,Vedral2008quantifying,Laflorencie2016quantum,Gogolin2016equilibration,Ueda2020quantum,Zeng2015quantum}.
Especially in the latter, entanglement has significant effects on the macroscopic behavior of quantum materials \cite{Ghosh2003entangled,Brukner2006crucial} and is intimately connected to the appearance of quantum spin liquidity \cite{Balents2010,Savary2016quantum,Wen2017,Semeghini2021}, topological order \cite{Zeng2015quantum,Wen2017}, quantum criticality \cite{Prochaska2020singular}, and strange metallic behavior \cite{Zaanen2019planckian,chen2019incoherent}. Therefore, efforts to identify and quantify quantum correlations in solids have the potential to reveal new collective phenomena and to advance our capability to harness their functionalities.
\par While small entangled systems can be experimentally characterized via tomography \cite{james2001measurement,Haffner2005scalable} or quantum interference methods \cite{Islam2015measuring}, diagnosing entanglement in solids requires alternative approaches linked to accessible experimental observables \cite{Amico2008entanglement,Vedral2008quantifying,guhne2009entanglement}. A possible strategy relies on determining the expectation values of operators called ``entanglement witnesses'' \cite{Coffman2000distributed,amico2004dynamics,Roscilde2004studying,Brukner2006crucial,Amico2006divergence,Pezz2009,Hyllus2012,Toth2014}.
The witness selection depends on the type of system and entanglement of interest, but a prominent quantity is the quantum Fisher information (QFI). At equilibrium, the QFI can be rigorously extracted from a sum-rule integral of the Kubo response function \cite{Hauke2016measuring} and acts as a multipartite entanglement witness if its value exceeds classical expectations. Its connection with experimentally accessible response functions motivated recent inelastic neutron scattering studies of multipartite entanglement in a variety of spin chains \cite{Mathew2020experimental,Laurell2021quantifying,Scheie2021witnessing} and defines a model-independent pathway to detect entangled states in quantum materials.

\par Witnessing entanglement with the QFI or other operators has significant implications well beyond the study of quantum systems at equilibrium.
Recently, ultrafast laser pulses have enabled new pathways to drive quantum materials through nonequilibrium phase transitions, or induce entirely new states of matter without apparent equilibrium analogues \cite{Hu2014optically,Kaiser2014optically,Mitrano2016possible,Buzzi2020photomolecular,Wang2013observation,Mahmood2016selective,McIver2020light,delatorre2021nonthermal}. Probing quantum entanglement for these dynamical phenomena is crucial in order to understand their microscopic origin and to identify the possible role of transient quantum coherence, especially in systems without obvious order parameters \cite{claassen2017dynamical,Kaneko2019photoinduced,peronaci2020enhancement,Li2020eta}. Hence, it becomes important to investigate how a time-dependent QFI reflects the evolution of entanglement through a nonequilibrium phase transition.

\par In this Letter, we show that the QFI is a robust witness of time-dependent multipartite entanglement across a prototypical nonequilibrium phase transition. We consider the experimentally relevant case of a fermion chain with a dynamically-tuned Coulomb repulsion \cite{singla2015THz,tancogne2018ultrafast,beaulieu2021ultrafast,baykusheva2022ultrafast} and map its entanglement dynamics for different driving conditions. Upon ramping the interaction strength, the system undergoes a quantum phase transition from a disordered to an ordered phase. The QFI witnesses an increase in multipartite entanglement while ramping the Coulomb interaction. For adiabatic ramps, the QFI exhibits a well-defined local maximum at the critical point, consistent with expectations for equilibrium quantum phase transitions \cite{Hauke2016measuring}. By contrast, in a diabatic regime its increase persists deep into the ordered phase with distinctive oscillatory behavior. Crucially, such enhancement is robust against the introduction of local decoherence processes, thus underscoring the possibility to witness entanglement dynamics of quantum systems coupled to realistic dissipative baths. Our results are immediately relevant to the nonequilibrium dynamics of one-dimensional Mott insulators in the strong coupling limit [e.g. [Ni(chxn)$_2$Br]Br$_2$, K-TCNQ, and (ET)-F$2$TCNQ] \cite{Iwai2003ultrafast,Okamoto2007photoinduced,Uemura2008ultrafast,Mitrano2014pressure,singla2015THz,Wall2011quantum,Sono2022phonon}, and, through a mapping onto an equivalent spin model, of spin chain systems such as Cs$_2$CoCl$_4$ \cite{Laurell2021quantifying}, KCuF$_3$ \cite{Scheie2021witnessing}, or  [Cu($\mu$-C$_2$O$_4$)(4-aminopyridine)$_2$(H$_2$O)]$_n$ \cite{Mathew2020experimental}. More broadly, our work defines a strategy to distinguish different dynamical regimes in driven materials via quantum information methods.

\textit{Model.}--- We consider a half-filled chain of spinless fermions interacting through a nearest-neighbor Coulomb repulsion. The model Hamiltonian reads,
\begin{equation}
\label{eq:Hamiltonian_fermionic}
    \Ham(t)=-\frac{J}{2}\sum_j
	(\cop_{j}^\dagger \cop_{j+1}+\mathrm{H.c.})+V(t)\sum_j\tilde{n}_{j}\tilde{n}_{j+1},
\end{equation}
where $\cop_{j}^\dagger$ ($\cop_{j}$) is a creation (annihilation) operator at site $j$, $\tilde{n}_{j}=\cop_{j}^{\dagger}\cop_{j}-1/2$ is the number operator relative to half filling, and $J$ is a constant hopping amplitude setting the energy scale of our model. We consider a time-dependent nearest-neighbor Coulomb interaction $V(t)$, which can be realized in ultrafast optical experiments through dielectric screening enhancement \cite{tancogne2018ultrafast}, coherent Floquet dressing \cite{Mentink2015,delatorre2021nonthermal}, or transient crystal lattice distortions \cite{Disa2021}. \textcolor{black}{For simplicity, we ramp up the interaction strength at constant velocity as $V(t) = Jvt$ for all the calculations presented in this work.} At equilibrium, this system exhibits a well-known quantum phase transition at $V=1$ from a gapless Luttinger liquid (LL) with short-range correlations to a charge density wave (CDW) with long-range correlations.  
\par
\begin{figure}[tbp]
	\centering
	\includegraphics[width=\columnwidth]{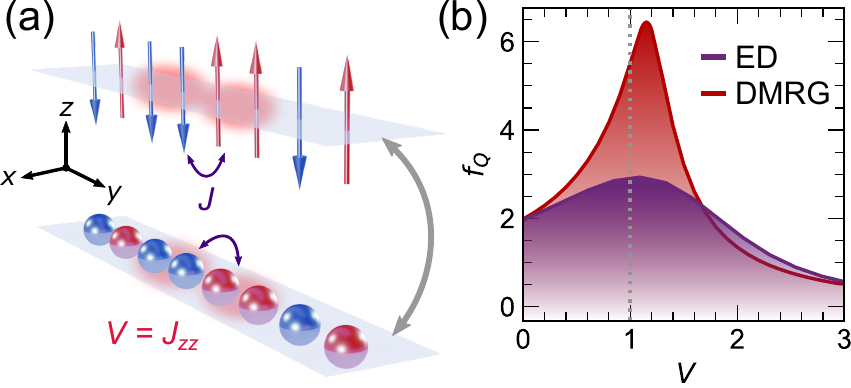}
	\caption{
	(a) Sketch of half-filled spinless fermion chain with nearest-neighbor Coulomb interaction [Eq.~\eqref{eq:Hamiltonian_fermionic}] (lower) and a corresponding spin-1/2 anisotropic Heisenberg model (upper). The fermionic hopping amplitude corresponds to the exchange coupling $J$ and the Hubbard interaction $V$ becomes the $z$-direction exchange coupling $J_{zz}$.
	(b) Equilibrium QFI density as a function of the intersite Coulomb repulsion $V$ computed using ED for a $L=10$ chain (purple) and using DMRG for an infinite chain (red), respectively. If $f_Q>1$, the system is at least bipartite entangled.
	}\label{fig:1_Equilibrium}
\end{figure}

 We map the spinless fermions onto an equivalent spin-1/2 anisotropic Heisenberg chain (Fig.~\ref{fig:1_Equilibrium}a) via a Jordan-Wigner transformation \cite{Giamarchi2004quantum,Altland2010Condensed} and add a small staggered magnetic field $h_z=0.005$ (cf. Sec.~\ref{sec:JW} and~\ref{sec:Finite-size} in the Supplementary Material (SM)~\cite{Supplementary}) in order to select one of the two classical N\'eel states. We have checked that our key results are robust against the specific value of the field. We study the nonequilibrium entanglement dynamics through exact diagonalization (ED) calculations using the Qu\-Spin \cite{weinberg2017quspin,weinberg2019quspin} and HPhi~\cite{kawamura2017} packages. In the equivalent spin formulation, the quantum phase transition occurs between the XY and antiferromagnetic phases. \textcolor{black}{
Here, we report ED calculations for a 10-site spin chain with periodic boundary conditions (PBC) and benchmark a selected subset of these against real-time density matrix renormalization group (DMRG) calculations~ \cite{Kennes_2018} on an infinite fermionic chain. ED calculations for chains of up to 24 sites (also with PBC) are included in the Supplementary Materials.}

\par We probe the entanglement dynamics by directly calculating the time-dependent QFI density using the instantaneous wavefunction describing the system. For a pure state $\psiG$, the QFI associated with an operator $\Oop_q$ is simply the connected correlation function~\cite{Hauke2016measuring}:
\begin{equation}\label{eq:QFI_definition_pure}
    \FQ = 4\Delta\left(\Oop_q\right)^2 = 4\left( \bra{\psiG}\Oop_q^2\ket{\psiG} - \bra{\psiG}\Oop_q\ket{\psiG}^2 \right).\textbf{}
\end{equation}
Since we focus on phase transitions to states with staggered correlations at wavevector $q=\pi$ \cite{Giamarchi2004quantum}, we choose the local generator $\Oop_\pi = \sum_l(-1)^l\hat{S}_l^z$ (formulated in the spin language). A value of the QFI density $f_Q\equiv \FQ/L>m$, where $m$ is a divisor of the system size $L$, signals that the state $\ket{\psiG}$ must be $m+1$-partite entangled \cite{Hyllus2012,Toth2012,Hauke2016measuring}.\\

\par \textit{Results.}--- Owing to the superposition of an increasing number of states due to quantum fluctuations, \textcolor{black}{a system approaching a non-topological quantum phase transition (i.e., one with a local order parameter on the quasi-ordered side of the phase transition) will exhibit enhanced multipartite entanglement and a QFI maximum around the critical point} \cite{Hauke2016measuring}. We first investigate the evolution of the QFI density $f_Q$ upon tuning the interactions across the equilibrium critical point. As shown in Fig. \ref{fig:1_Equilibrium}b, $f_Q$ correctly identifies the quantum phase transition ($V=1$) via a clear local maximum. This behavior is common to both the small and infinite-size limits (cf. Sec.~\ref{sec:Finite-size} in \cite{Supplementary}), whereby the chain length  mainly determines the sharpness of the QFI maximum with minor effects on its exact position.
In the gapless (LL) phase, the QFI density exceeds the classical bound $f_Q=1$, thus witnessing at least bipartite entanglement.  
\textcolor{black}{In the gapped (CDW) regime, the QFI density becomes featureless and decreases below the classical bound, consistent with the absence of multipartite entanglement deep inside the ordered phase.}

\begin{figure}[tbp]
	\centering
	\includegraphics[width=\columnwidth]{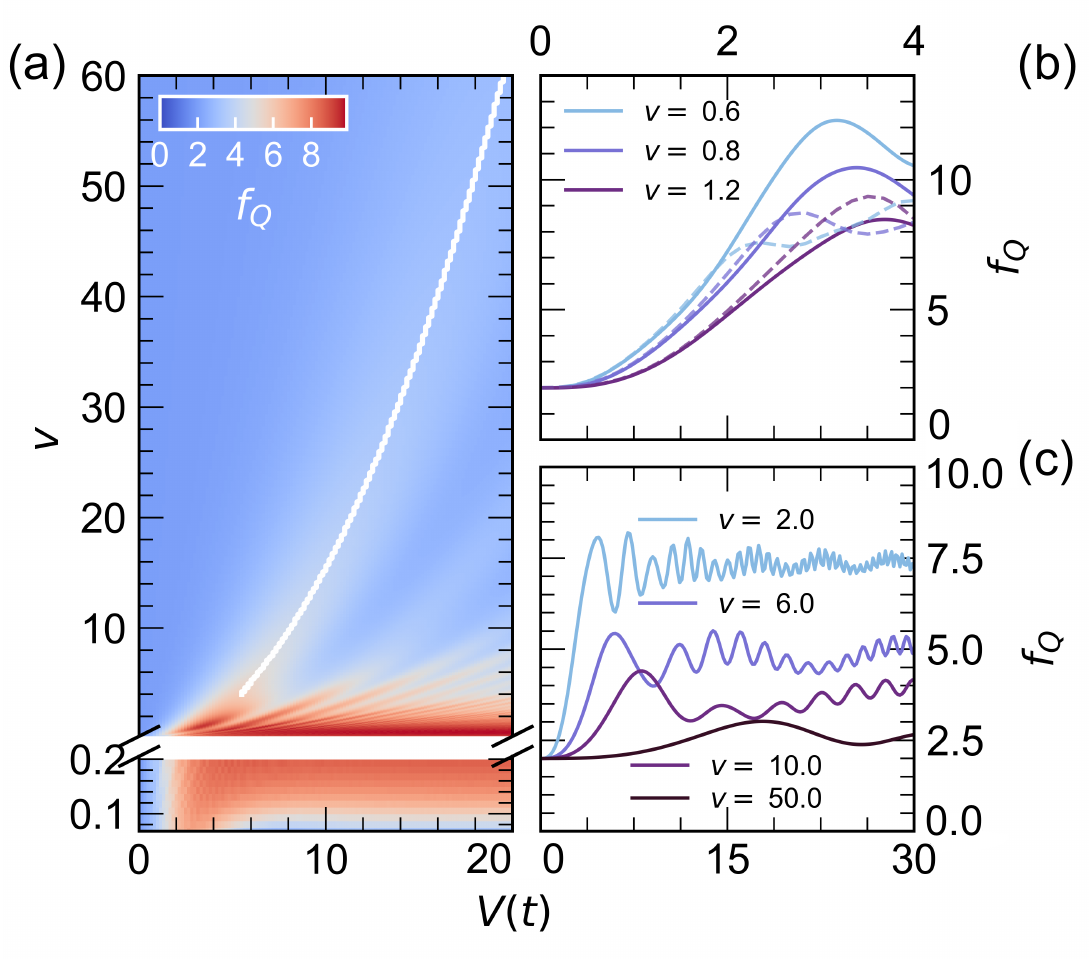}
	\caption{ (a) Nonequilibrium QFI density $f_Q=\FQ/L$ as a function of \textcolor{black}{the time-dependent interaction strength} and ramp velocity for $L=10$ sites. The white line marks the position of the instantaneous maxima of $f_Q$ in the impulsive region (see text). The lower part of the panel is a zoom of the quasi-adiabatic region. (b) Comparison between ED (dashed line) and DMRG (solid line) calculations of the time-dependent QFI density $f_Q$ for selected ramp velocities. (c) Time-dependent QFI density $f_Q$ for various ramp velocities. 
	}\label{fig:2_Dynamical}
\end{figure}
  
\par We now probe the nonequilibrium entanglement dynamics of our system. We consider a purely unitary evolution while ramping the interaction strength (see Fig.~\ref{fig:2_Dynamical}a). Starting from the noninteracting limit, we calculate the QFI of the time-evolved \textcolor{black}{initial} state $\ket{\psiG(t)} = \mathcal{T}\exp\left[-i\int_{t_0}^t \Ham(t')\mathrm dt'\right]\ket{\psiG(t_0)}$ for a fine mesh of ramp velocities. In the adiabatic limit ($v\rightarrow0$) the QFI density still peaks around the equilibrium quantum critical point. However, its nonequilibrium behavior exhibits significant differences. Upon increasing ramp speed, the local QFI maximum undergoes a long-lived enhancement with no signs of decay at large interaction strength. The transition to a diabatic regime occurs at relatively low velocity through a highly entangled state with $f_Q\sim L$. By decomposing the eigenstates of the instantaneous Hamiltonian into Fock states specifying the spins at each lattice site, we see that the entanglement growth of this region is mainly driven by a ``Schr\"{o}dinger-cat''-like superposition of nearly-degenerate N\'{e}el states $|\uparrow\downarrow\uparrow\downarrow\ldots\rangle$ and $|\downarrow\uparrow\downarrow\uparrow\ldots\rangle$~\cite{Greenberger1989, Pezze2018, Ekert1998, Neumann2008, Bouwmeester1999, Pan2001, Wang2018, Leibfried2005, Monz2011, Friis2018, DiCarlo2010, Song2017, Vlastakis2013, Omran2019}. \textcolor{black}{As we show later, this coherent superposition is not stable with respect to decoherence.} At higher ramp velocity ($v> 3.5$), the nonequilibrium enhancement of the QFI density peaks at larger interaction strengths (earlier times) and at a progressively lower level, thus suggesting a lower degree of multipartite entanglement. Intriguingly, the \textcolor{black}{existence of the main peak of the time-dependent QFI density} does not depend on the system size (see Fig.~\ref{fig:2_Dynamical}b) while its position is insensitive to the presence of additional interaction terms in the equilibrium model (Sec.~\ref{sec:Next_nearest_neighbor} in \cite{Supplementary}). Hence, it represents a genuine nonequilibrium feature which does not extrapolate to the equilibrium critical point~\cite{tsuji2013nonthermal} and can be used to define a critical speed $v^\ast$ separating adiabatic and diabatic ('impulsive') dynamical regimes. In addition, the QFI density enhancement is accompanied by a characteristic oscillatory behavior (see Fig.~\ref{fig:2_Dynamical}c). These oscillations contain both size-dependent and size-independent frequency components, as analyzed in greater detail in Sec.~\ref{sec:Time_Frequency} of the SM~\cite{Supplementary}. Through a time-frequency analysis via a sliding-window Fourier transform, we assign these oscillations to transitions across the gap of the instantaneous Hamiltonian $\omega(t)\approx\Delta(t)$ \cite{Pollmann2010dynamics,Canovi2014dynamics}.

\begin{figure}[tbp]
	\centering
	\includegraphics[width=\columnwidth]{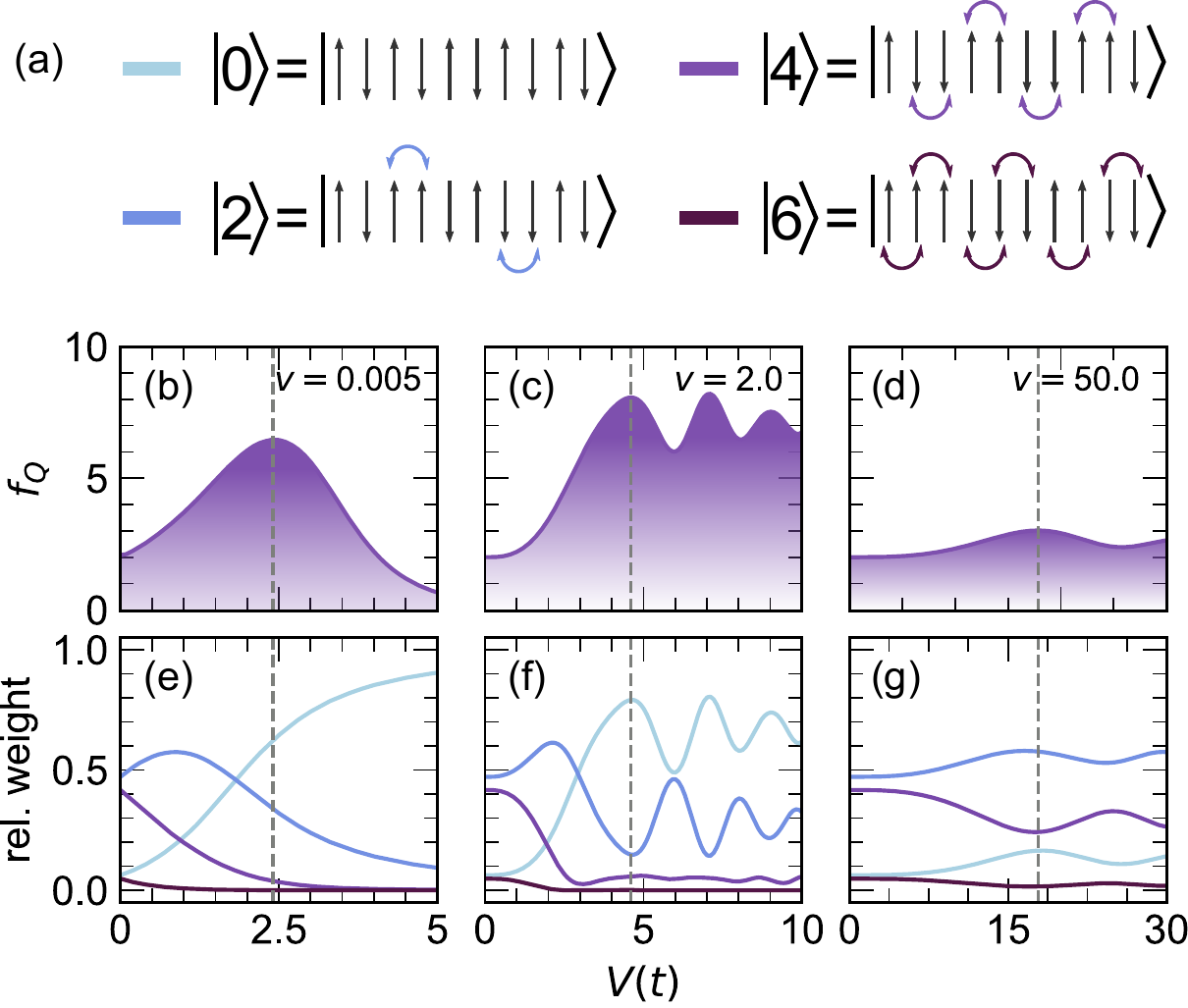}
	\caption{Nonequilibrium QFI (top) and time-dependent density of defects (bottom) for selected velocities, \textcolor{black}{(b),(e) $v=0.005$, (c),(f) $v=2$, and (d),(g) $v=50$}. Dashed vertical lines indicate the position of the first QFI maximum. \textcolor{black}{(a)} Chain configurations with a variable number of domain walls. The density of defects is given by the sum of the weights of time-evolved states containing zero to three domain walls. \textcolor{black}{All calculations have been performed using the ED method.}}\label{fig:4_Defects}
\end{figure}

\par To microscopically interpret the observed nonequilibrium entanglement dynamics, we decompose the instantaneous wavefunction $\psiG(t)$ as a weighted superposition of N\'{e}el-like spin configurations with increasing number of domain walls (or \textit{defects}, s. Fig.~\ref{fig:4_Defects}a). We then map the nonequilibrium QFI density at selected ramp velocities onto the time-dependent behavior of these weights (see Fig.~\ref{fig:4_Defects}). In the adiabatic limit ($v\lesssim 0.03$, see Fig.~\ref{fig:4_Defects}b, f), our system at $t_0=0$ is initialized in a disordered phase with a high density of defects. As the interaction strength grows, the time-evolved wavefunction $\psiG(t)$ follows the ground state of the equilibrium Hamiltonian with effective interaction $V(t)$ and shows an increased superposition of pure N\'{e}el states without domain walls. Since the equally-weighted superposition of N\'{e}el states is maximally entangled and the underlying wavefunction cannot be represented as a product state, the QFI density $f_Q$ increases to a value close to the system size ($f_Q\sim L$). Then, above the transition into the ordered phase, $f_Q$ decays as the system selects one of the two N\'{e}el configurations and reduces the amount of multipartite entanglement.
\par At small ramp velocities ($0.175 < v < 3.5$, see Fig.~\ref{fig:4_Defects}c, f), the entanglement dynamics becomes diabatic and involves the excitation of states with non-zero defect density. The system still features a growth of the superposition between pure N\'{e}el configurations with the QFI density $f_Q$ approaching the maximum value allowed by the system size. However, the departure from adiabaticity leads to the excitation of multiple low-lying states with a finite number of domain walls, their population being periodically redistributed as a function of time. 
At large ramp velocities ($v>3.5$, see Fig.~\ref{fig:4_Defects}d, g), the QFI dynamics is deep in the diabatic regime. The time-evolved wavefunction does not follow the ground state evolution and its composition becomes skewed towards states with a non-zero number of domain walls. The superposition of pure N\'{e}el configurations is suppressed and the overall entanglement content of our model is reduced. In other words, a rapid change of the interaction strength drives the system into states with a larger number of defects and lower multipartite entanglement.

Since the QFI behavior is closely related to the proliferation of defects, we verify whether this entanglement witness follows a dynamics of the Kibble-Zurek (KZ) type \cite{Kibble1976, Zurek1985}. In the KZ picture, the dynamics of a system undergoing a quantum phase transition, including the scaling of the density of defects and other observables as a function of the quench velocity, is determined by its behaviour at the critical point \cite{Polkovnikov2005,Zurek2005,Dziarmaga2005}. While the validity of the KZ mechanism in the quantum regime has been established for trajectories across a critical point, our quench from the disordered to the ordered phase requires crossing an entire critical region ($0\le\Delta<1$)~\cite{Schuetzhold2006, Pellegrini2008} and the critical point at $V=1$ is of the Berezinskii-Kosterlitz-Thouless (BKT) type~\cite{Dalmonte2015}. \textcolor{black}{Figure~\ref{fig:4_Defects} and the corresponding analysis strongly imply the existence of a tight connection between the extrema of the dynamical QFI and  the  number of defects as a function of time. In Fig.~\ref{fig:Defects_Extrema} of the SM, we show that the maximum of $f_Q (t)$ in fact coincides with a minimum of the integrated density of defects, consistent with the fact that the predominance of the N\'{e}el-like configuration is conducive to higher entanglement levels. Therefore, in the following we interpret the scaling of the time instant $t^\ast$ maximizing the dynamical QFI density $f_Q$ in terms of the KZ paradigm.} In the impulsive regime, we find that the nearest-neighbour Coulomb repulsion $V(t^\ast)$ maximizing the QFI density $f_Q$ is consistent with the KZ expectation and follows a dynamical scaling given by the power law $V(t^\ast)\propto v^{\alpha}$ with $\alpha \approx 1/2$ \textcolor{black}{($0.5295\pm0.0009$)}. In Sec.~\ref{sec:Limits} of the SM~\cite{Supplementary}, we relate $\alpha$ with the critical exponent of the system and the scaling of the gap at the critical point $V =1$. Remarkably, for the systems sizes investigated here the KZ scaling of $V(t^\ast)$ is insensitive to diverse microscopic perturbations of the equilibrium model such as second-neighbor interactions, decoherence, or external fields \cite{Supplementary}. 

\begin{figure}[tbp]
	\centering
	\includegraphics[width=\columnwidth]{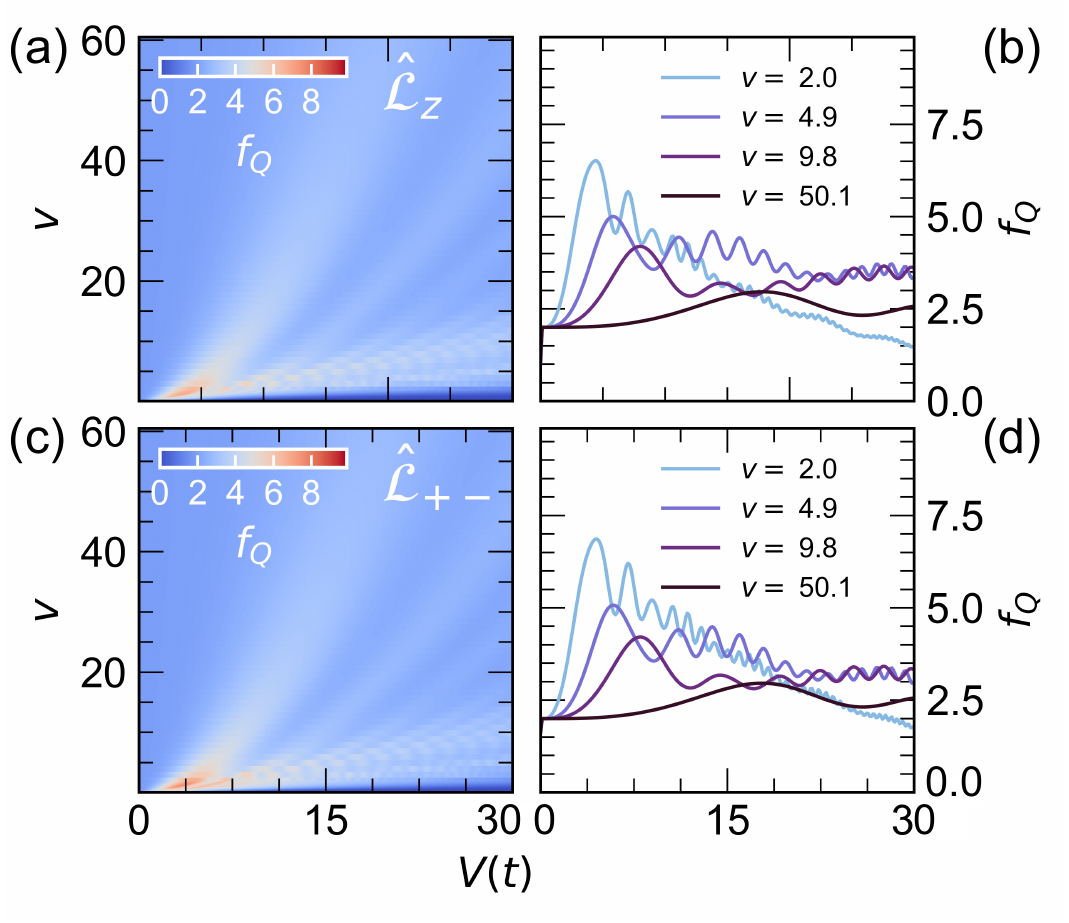}
	\caption{Nonequilibrium QFI density $f_Q=\FQ/L$ as a function of time and ramp velocity for $L=10$ sites and at fixed decoherence rate $\gamma=0.01$. (a),(c) Evolution of $f_Q$ in the presence of the Lindbladian jump operators $\Lb_{z}$ and $\Lb_{+-}$, respectively. (b),(d) Time-dependent $f_Q$ curves for selected ramp velocities. \textcolor{black}{The calculations have been performed employing the ED method.}
	}\label{fig:3_Decoherence}
\end{figure}

Since most nonequilibrium experiments on condensed matter systems involve the coupling to an external environment, we now verify that the QFI density enhancement survives the presence of quantum dissipation. We evolve the density matrix $\rhoDM(t) = \ket{\psiG(t)}\bra{\psiG(t)}$ according to a Lindblad master equation with a decoherence rate $\gamma$
\begin{equation}
\label{eq:Lindblad}
\resizebox{.99\hsize}{!}{$\dot{\rhoDM}(t) = -i\left[\Ham(t),\rhoDM(t)\right] + 2\gamma \sum_{l=0}^{L-1} \left(\Lb_l\rhoDM(t)\Lb_l^\dagger - \frac{1}{2}\left\{ \Lb_l^\dagger\Lb_l,\rhoDM(t)\right\} \right)$}.
\end{equation}
We choose two different quantum jump operators $\Lb_l$, namely $\Lb_{z}\equiv \hat{\sigma}_l^z$ and $\Lb_{+-}\equiv  \hat{\sigma}_l^+\hat{\sigma}_{l+1}^-$. The former describes local dephasing, relevant to the description of decay processes involving local degrees of freedom, such as molecular vibrations \cite{Mitrano2014pressure}. The latter encodes instead nonlocal quantum decoherence \cite{Eisler2011crossover} relevant to systems featuring magneto-elastic coupling where the spin-exchange process is coupled to a bath of vibrational oscillators. \textcolor{black}{ The calculation of the QFI in this case follows the definition for a mixed state}
\begin{equation}
   \textcolor{black}{ \FQ(t) = 2 \sum_{i,j}\frac{(\epsilon^i_t-\epsilon^j_t)^2}{\epsilon^i_t+\epsilon^j_t} 
    \left|\bra{\varphi^i_t}\Oop_\pi\ket{\varphi^j_t}\right|^2 } \label{eq:QFI_DM_main},
\end{equation}
\textcolor{black}{where $\epsilon^i_t$ and $\ket{\varphi^i_t}$ are eigenvalues and eigenvectors of $\rhoDM$ at time $t$ (see Sec.~\ref{sec:Lindblad} in the SM~\cite{Supplementary} for further details)}.  
As shown in Fig. \ref{fig:3_Decoherence}, the decoherence terms suppress the ``Schr\"{o}dinger-cat''-like state at low ramp speeds and introduce a decay of the QFI density for increasing interaction strength. However, the main nonequilibrium QFI peak is preserved and persists up to coupling rates of $\gamma\approx 0.1$ \cite{Supplementary} with only minor changes in location and magnitude. A decoherence rate $\gamma=0.1$ is consistent with the experimental quasiparticle recombination rates of certain quasi-1D Mott insulators \cite{Mitrano2014pressure}. Hence, it should be possible to use the QFI density to witness nonequilibrium entanglement dynamics in realistic quantum systems subject to decoherence.\\

\textit{Conclusion.}--- We have investigated the nonequilibrium evolution of a driven fermion chain in terms of its many-body entanglement properties. By dynamically tuning the intersite Coulomb repulsion, we witness the multipartite entanglement dynamics with the time-dependent QFI across a prototypical quantum phase transition and find it robust against the introduction of decoherence.
\par The existence of a rigorous connection between the QFI and the dynamical response of quantum systems at equilibrium has motivated recent attempts to witness multipartite entanglement in inelastic neutron scattering experiments \cite{Laurell2021quantifying,Mathew2020experimental,Scheie2021witnessing}. The same protocols could be extended to certify the presence of entangled states in driven systems with ultrafast resonant inelastic x-ray scattering and optical methods. Our work establishes the QFI as a valid nonequilibrium entanglement witness \textcolor{black}{for driven condensed matter systems beyond its current use in quantum simulators and quantum information science \cite{Smith2016manybody,Almeida2021from,Yu2022critical,Zhong2013fisher,Guo2020distributed,xu2022metrological}}. 
However, further progress \textcolor{black}{in this direction} requires mapping the QFI operator to transient response functions (thus abandoning the fluctuation-dissipation theorem as the key tenet of Ref.~\cite{Hauke2016measuring}), determining appropriate quantum bounds for the relevant observables, and
\textcolor{black}{controlling sources of experimental uncertainties (e.g. energy resolution broadening or incorrect normalization of experimental spectra) which could lead to an inaccurate QFI extraction \cite{Hauke2016measuring,Scheie2021witnessing,Laurell2021quantifying}}.
These steps will allow extending the use of entanglement witnesses to future nonequilibrium spectroscopy experiments on quantum materials.
\par Our findings pave the way towards a deeper understanding of the role of quantum correlations in photoinduced phase transitions. The entanglement dynamics studied here is immediately relevant to laser-driven quasi-one-dimensional Mott insulators and spin chain materials~\cite{Kennes_2018} and represents a template to understand the more general behavior of correlated quantum materials out of equilibrium. In particular, entanglement correlations might allow for a better characterization of hitherto incompletely understood pathways towards hidden metastable phases \cite{vedral2004high,Li2020eta}, the identification of dynamical quantum phases without obvious order parameters~\cite{Jiang2012qsl, Szasz2020qsl}, and could be key in unraveling the mystery behind light-induced phases of matter without equilibrium analogues \cite{Tindall2020dynamical}.

We would like to thank S.~R.~Clark, J.~Marino, Y.~Wang, N.~Yao, and P.~Zoller for insightful discussions. \textcolor{black}{This work was supported by the U.S. Department of Energy, Office of Basic Energy Sciences, Early Career Award Program, under Award No. DE-SC0022883. M. M. further acknowledges support by the Aramont Fellowship Fund for Emerging Science Research at Harvard University.} D.~R.~B. was supported by the Swiss National Science Foundation through Project No. P400P2\_194343. M.~C. acknowledges support by NSF Grant No. DMR-2132591. D.~M.~K. acknowledges funding by the Deutsche Forschungsgemeinschaft (DFG, German Research Foundation) via Germany's Excellence Strategy – Cluster of Excellence Matter and Light for Quantum Computing (ML4Q) EXC 2004/1 – 390534769 and within the RTG 1995. M.~A.~S. acknowledges financial support through the Deutsche Forschungsgemeinschaft (DFG, German Research Foundation) via the Emmy Noether program (SE 2558/2). We also acknowledge support from the Max Planck-New York City Center for Non-Equilibrium Quantum Phenomena.

%

\end{document}


\newcommand{\su}{\uparrow}
\newcommand{\sdn}{\downarrow}
\newcommand{\HXXZ}{\hat{\mathcal{H}}_{XXZ}}
\newcommand{\Jeff}{J^z_\mathrm{eff}}
\newcommand{\psiG}{{\psi}_0}
\newcommand{\Oop}{\hat{\mathcal{O}}}
\newcommand{\FQ}{\mathcal{F}_Q}
\newcommand{\rhoDM}{\hat{\rho}}
\newcommand{\Lb}{\hat{\mathcal{L}}}
\newcommand{\Ham}{\hat{\mathcal{H}}}
\newcommand{\fbeat}{\nu_\mathrm{osc}}
\newcommand{\txt}{t^\ast}
\newcommand{\Jxt}{V(t^\ast)}
\newcommand{\vcrf}{v^{\ast,1}}
\newcommand{\vcrs}{v^{\ast,2}}
\newcommand{\cop}{\hat{c}}
\newcommand{\cMstagg}{\mathcal{M}_\mathrm{stagg}}
\renewcommand{\thepage}{S\arabic{page}}
\renewcommand{\thesection}{S\arabic{section}}
\renewcommand{\thetable}{S\arabic{table}}
\renewcommand{\thefigure}{S\arabic{figure}}

\title{Supplementary Material for: \\
Witnessing Nonequilibrium Entanglement Dynamics in a Quenched Quantum Chain}

\newcommand{\affiliationHarvard}{
Department of Physics, Harvard University, Cambridge, Massachusetts 02138, USA
}

\newcommand{\affiliationMPSD}{
Max Planck Institute for the Structure and Dynamics of Matter, Center for Free-Electron Laser Science (CFEL),
Luruper Chaussee 149, 22761 Hamburg, Germany
}

\newcommand{\affiliationRWTH}{
Institut f\"ur Theorie der Statistischen Physik, RWTH Aachen University, 52056 Aachen, Germany and JARA-Fundamentals of Future Information Technology, 52056 Aachen, Germany
}

\newcommand{\affiliationPenn}{Department of Physics and Astronomy, University of Pennsylvania, Philadelphia, PA 19104, USA}

\author{Denitsa R. Baykusheva}
\affiliation{\affiliationHarvard}

\author{Mona H. Kalthoff}
\affiliation{\affiliationMPSD}

\author{Damian Hofmann}
\affiliation{\affiliationMPSD}

\author{Martin Claassen}
\affiliation{\affiliationPenn}

\author{Dante M. Kennes}
\affiliation{\affiliationRWTH}
\affiliation{\affiliationMPSD}

\author{Michael A. Sentef}
\affiliation{\affiliationMPSD}

\author{Matteo Mitrano}
\affiliation{\affiliationHarvard}

\date{\today}
\maketitle

\section{Model system}\label{sec:JW}

In this work, we map a half-filled chain of spinless fermions interacting through nearest neighbor Coulomb repulsion onto a corresponding spin Hamiltonian. The initial model Hamiltonian is given by:
\begin{equation}
\label{eq:Hamiltonian_fermionic}
    \Ham(t)=-\frac{J}{2}\sum_j
	(\cop_{j}^\dagger \cop_{j+1}+\mathrm{H.c.})+V(t)\sum_j\tilde{n}_{j}\tilde{n}_{j+1},
\end{equation}
where $\cop_{j}^\dagger$ ($\cop_{j}$) is a fermionic creation (annihilation) operator at site $j$, $\tilde{n}_{j}=\cop_{j}^{\dagger}\cop_{j}-1/2$ is the number operator relative to half filling, $J$ is a constant hopping amplitude, and $V(t)$ is a time-dependent, nearest-neighbor Coulomb interaction.
Through the Jordan-Wigner transformation \cite{Giamarchi2004quantum}, this charge Hamiltonian maps onto an equivalent spin-1/2 anisotropic Heisenberg (XXZ) chain (Fig.~1a in the main text):
\begin{equation}\label{eq:Hamiltonian_Spin}
	\Ham_{\mathrm{sp}}(t)
	=\sum_j\left[-\frac{J}{2}\left(\hat{S}_{j}^{+}  \hat{S}_{j+1}^{-}+\mathrm{H.c.}\right) + V(t)\hat{S}_{j}^{z}\hat{S}_{j+1}^{z}\right],
\end{equation}
where $\hat{S}_{j}^{\pm}= \frac{1}{2} \left(\hat{S}_{j}^x\pm \hat{S}_{j}^y\right)$ and $\hat{S}_{j}^\alpha$ ($\alpha={x,y,z})$ are the usual spin operators defined in terms of the Pauli matrices $\hat{S}^\alpha = \tfrac{1}{2}\hat{\sigma}_\alpha$. In this picture, the hopping amplitude $J$ becomes the exchange coupling while $\Delta(t) =  V(t) / J$ quantifies the anisotropy of the spin interactions.
\par At equilibrium, this dual quantum chain exhibits well-known quantum phase transitions. Upon increasing $\Delta=V/J$, the fermionic chain evolves from a gapless Luttinger liquid (LL) phase with short-range correlations to a charge density wave (CDW) phase with long-range correlations. The XXZ chain instead undergoes two separate transitions into an Ising ferromagnet ($\Delta<-1$) and antiferromagnet ($\Delta>1$), while for ($|\Delta|<1$) it exhibits an XY phase \cite{Giamarchi2004quantum}. 
Since the dual charge and spin formulations are one-to-one equivalent, we choose to study the time-dependent dynamics of the quantum spin chain using exact diagonalization (ED). The ED calculations for finite-size chains extending up to $L=24$ sites are performed using the Qu\-Spin \cite{weinberg2017quspin,weinberg2019quspin} as well as the HPhi~\cite{kawamura2017} packages. 
The antiferromagnetic XXZ ground state in a finite-size system contains a mixture of nearly-degenerate states (notably $\ket{\su\sdn\ldots\su\sdn}\pm\ket{\sdn\su\ldots\sdn\su}$). In order to break this near degeneracy, which turns into complete degeneracy in the thermodynamic limit, we also introduce a small staggered magnetic field $\Ham_{\mathrm{ext}}=\sum_j(-1)^jh_z\hat{S}_{j}^{z}$ to select a specific spin configuration (cf. next section). This ensures that the disconnected part of the spin correlation function in Eq.~(2) in the main text vanishes in the limit $\Delta\rightarrow\infty$. Our spin sector quench dynamics is then benchmarked for selected conditions against real-time density matrix renormalization group (DMRG) calculations \cite{Kennes_2018} of the time evolution of an infinite chain. 

\section{Finite-size effects on the Quantum Fisher Information calculations}\label{sec:Finite-size}
\begin{figure}[h]
	\centering
	\includegraphics[scale = 1]{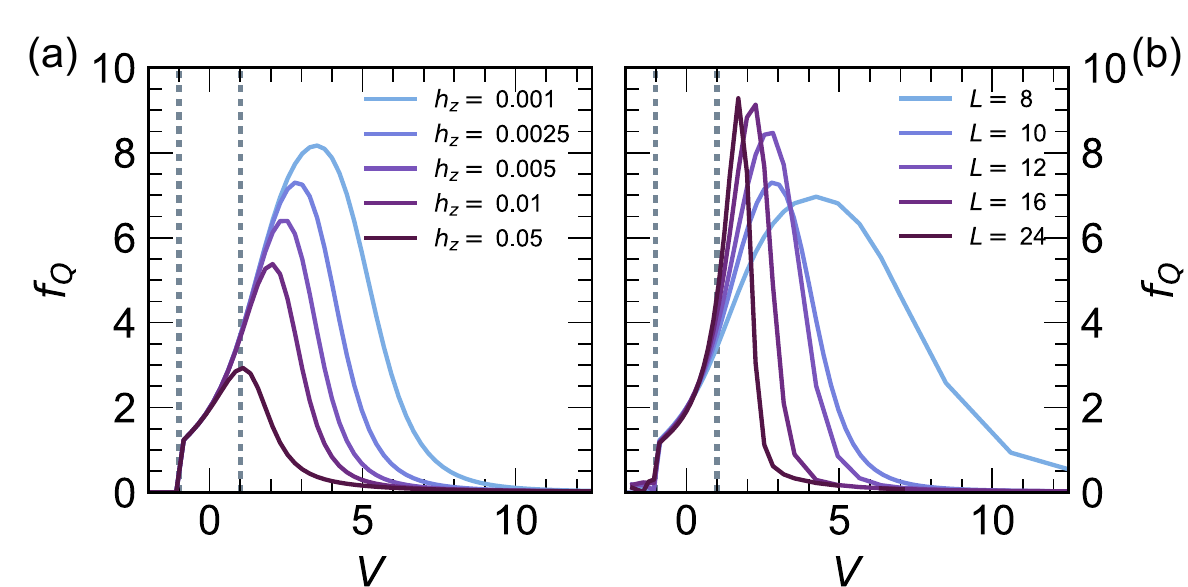}
	\caption{(a) Equilibrium QFI density $f_Q$ for a spin chain with $L=10$ for  different values of the staggered magnetic field $h_z$. (b) QFI density as a function of the chain length for a fixed $h_z =1.414$.}\label{fig:Finite_Size}
\end{figure}

In the antiferromagnetic (AFM) Ising limit $\Delta\rightarrow\infty$, the ground state of the XXZ chain in the spin sector is degenerate in the thermodynamic limit ($L\rightarrow\infty$) and is given by the two possible N\'{e}el states $\ket{\su\sdn\su\sdn\ldots}$ and $\ket{\sdn\su\sdn\su\ldots}$, related to each other via a site translation. At finite $L$, the two lowest-energy spin configurations $\ket{\su\sdn\su\sdn\ldots}\pm\ket{\sdn\su\sdn\su\ldots}$ are split by a finite energy gap that decreases as $\propto e^{-\alpha L}$ with increasing system size (with $\alpha>0$). For the AFM ground state at finite $L$, the disconnected part of the spin correlation function in Eq.~(2) of the main text, i.e.\ $\sum_{l,k}(-1)^{l+k}\left\langle\hat{S}^z_l\right\rangle\left\langle\hat{S}^z_{k}\right\rangle$, vanishes, leading to an asymptotically finite value of the QFI $F_Q$. In order to reinstate the proper behaviour of the QFI certified by the DMRG calculations in the charge sector, we introduce a staggered degeneracy-breaking magnetic field term $\Ham_{\mathrm{ext}}=\sum_j(-1)^jh_z\hat{S}_{j}^{z}$ to select a specific N\'{e}el component of the ground state and restore the proper limit of $F_Q$. Consequently, the position of the maximum of the equilibrium QFI density shown in Fig.~1b of the main text thus becomes dependent on the magnitude $h_z$ of this additional term. In addition, both the magnitude of the QFI maximum as well as its position are dependent on the chain length $L$. In Fig.~\ref{fig:Finite_Size}, we plot the equilibrium QFI density in the spin sector as a function of the staggered magnetic field for a fixed chain length (panel~a) and as a function of the system size for a fixed $h_z$.

\section{Time-frequency analysis of the non-equilibrium QFI}\label{sec:Time_Frequency}
In Fig.~2 of the main text and the accompanying discussion, we outline the main features of the nonequilibrium QFI dynamics. Upon crossing the critical point $\Delta=1$, the driven system passes through a more  entangled state compared to the initial condition. Depending on the ramp velocity, the QFI density $f_Q(t) = F_Q(t)/L$ either maps onto the equilibrium phase diagram, thereby decaying asymptotically to zero as $t\rightarrow\infty$, or exhibits a rich oscillatory behaviour when the driving protocol reaches the impulsive limit at sufficiently high velocities. In this section, we focus on origin of the oscillatory dynamics, while the transition between adiabatic and impulsive regimes is discussed in Sec.~\ref{sec:Limits}.\\

\par
\begin{figure}[h]
	\centering
	\includegraphics[scale = 1]{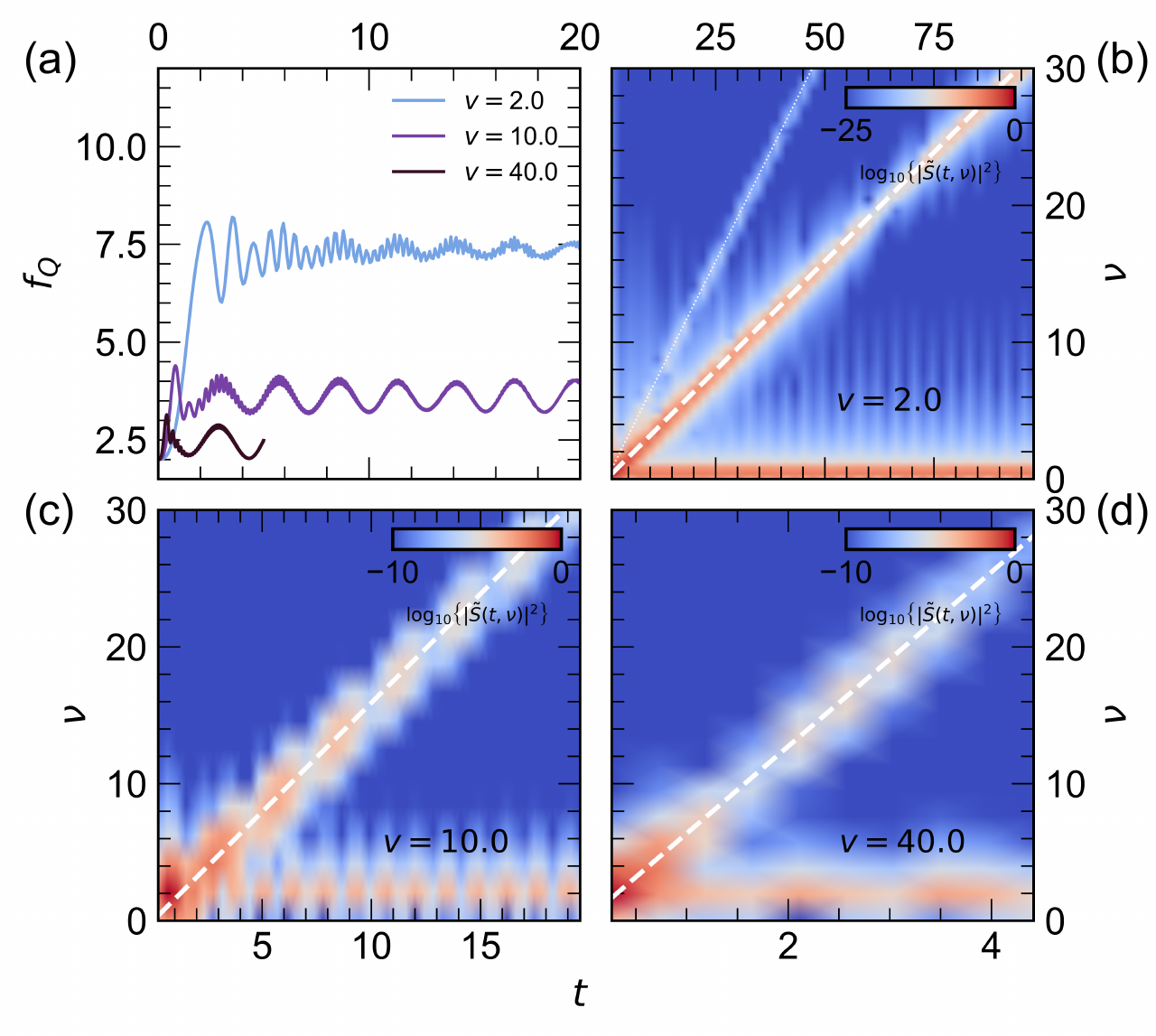}
	\caption{ (a) Linecuts from \textcolor{black}{Fig.~2c} in the main text, plotted as a function of time. (b) - (d) spectrograms of the time-dependent QFI density for three selected velocities: $v=2$ (b), $v=10$ (c), and $v=40$ (d). Note that the color scale is logarithmic. \textcolor{black}{Thick dashed lines indicate the dominant chirp mode $\overline{b}\sim \tfrac{v}{2\pi}$, while a thin dotted line indicates a second chirp mode with a slope of $2\overline{b}$ for $v=2$.}   }\label{fig:Time_Frequency}
\end{figure}

  We start by focusing our attention on the 1D linecuts of $f_Q(t)$ plotted in Fig.~2c of the main text. In panel~a of Fig.~\ref{fig:Time_Frequency}, we present three of these curves as a function of physical time instead of the time-dependent nearest-neighbour Coulomb repulsion $V(t)$. Both representations are related through the transformation $t=V(t)/v$. The time-domain representation allows us to better appreciate characteristic dynamical features of the non-adiabatic regime (\textcolor{black}{$v>3.5$}). First, all curves in Fig.~\ref{fig:Time_Frequency}a reveal a slow oscillation at $\fbeat\sim\textcolor{black}{0.36}$, independent of the driving speed. Furthermore, all curves also contain very fast oscillations, evident at early times and featuring a pronounced chirp. In order to isolate the different frequency contributions and characterize the change of the non-stationary contributions as a function of time, we perform a short-time Fourier transform (STFT) analysis. In the resulting spectrogram, a stationary frequency manifests itself as a horizontal line, whereas a linear change of the frequency over time corresponds to a linear chirp. \\

\par In panels b-d of Fig.~\ref{fig:Time_Frequency}, we plot the spectrograms corresponding to the three  velocities $v=2.0$, $v=10.0$, and $v=40.0$. The $v$-independent oscillations correspond to the low-frequency mode at $\fbeat \sim \textcolor{black}{2\pi\Delta_{01} \approx 0.36}$. The fast oscillations show a linear chirp, \textcolor{black}{which emerges outside of the adiabatic limit}, with a slope given by $\textcolor{black}{\overline{b}\sim \tfrac{v}{2\pi}}$.\\

\par
\begin{figure}[h]
	\centering
	\includegraphics[scale = 0.75]{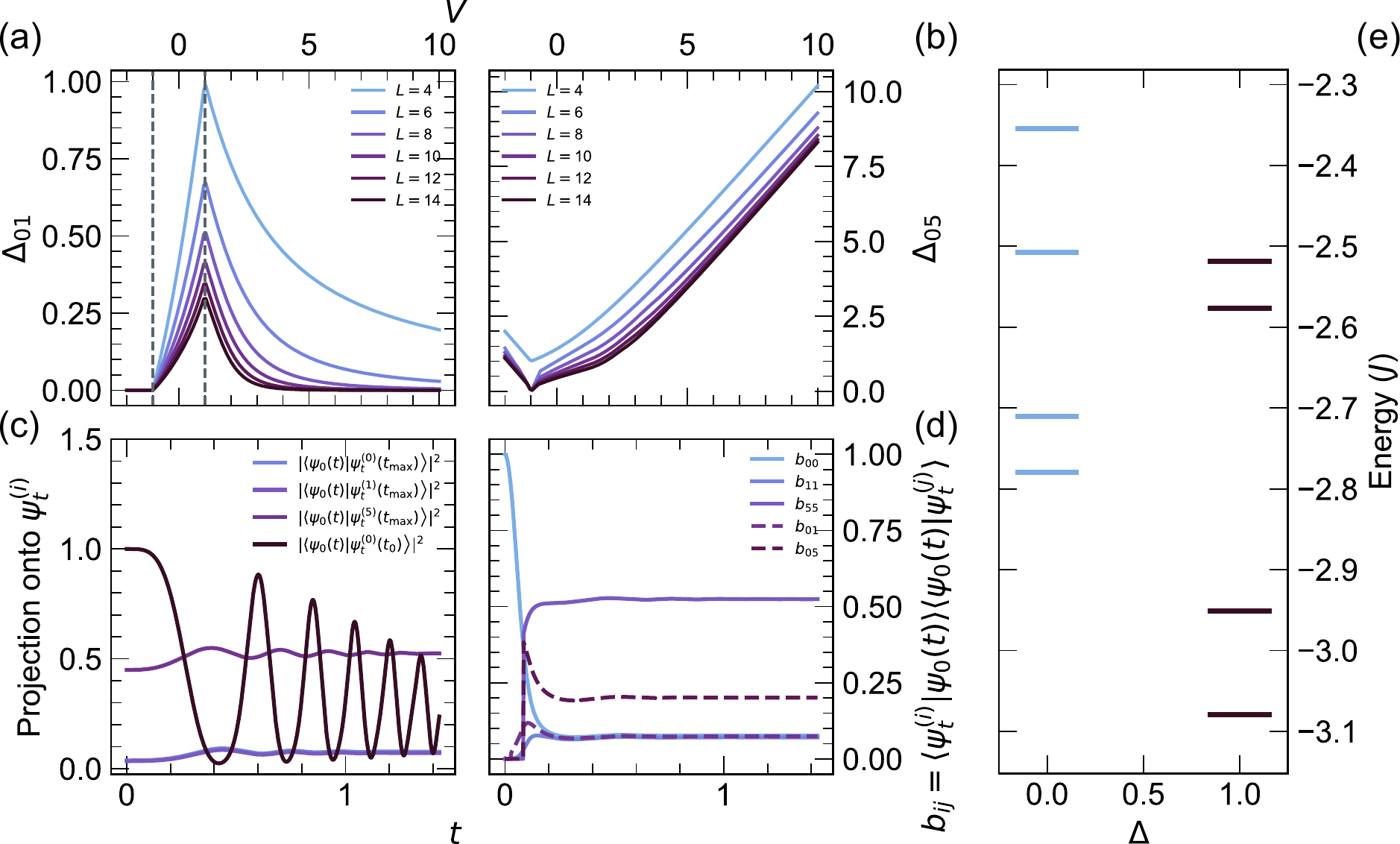}
	\caption{(a)-(b) Energy separations between the ground state and the first (a) resp.\ the fifth (b) excited states of the chain, plotted as a function of the nearest-neighbour Coulomb repulsion for different chain lengths. A staggered magnetic field value of $h_z=0.005$ has been used. Grey vertical lines in (a) indicate the positions of the critical points. (c) Projection of the time-evolved state $\psiG(t)$ onto selected eigenstates of the \textit{time-dependent} Hamiltonian at the end of the time-evolution $\Ham(t=t_\mathrm{max})$. The dark-purple curve shows the projection of $\psiG$ onto the initial state at $t_0$. Panel (d) Time-dependence of selected matrix elements $b_{ij}$, given by products of overlaps of the time-propagated $\psiG(t)$ with the eigenstates of the \textit{instantaneous} Hamiltonian  $\Ham(t)$ at time $t$. Panel (e) energy diagram showing the relevant energy levels dominating the dynamics. Some of the energy levels are degenerate.}\label{fig:Projections}
\end{figure}

 We now discuss the microscopic origin of these features with the aid of Fig.~\ref{fig:Projections}. By analyzing the energy level diagram of the $L=10$ chain (with $h_z$ fixed at $0.005$), we can attribute the $\fbeat\sim \textcolor{black}{0.36}$ oscillation to a transition between the ground and the first excited level. The corresponding energy separation $\Delta_{01}$ is shown as a function of the nearest-neighbour Coulomb repulsion for different chain sized in Fig.~\ref{fig:Projections}a. For $L=10$ and sufficiently large anisotropies ($\Delta>3$), the gap size assumes an asymtoptic value of $\textcolor{black}{\tfrac{0.314}{2\pi}}$, hence, the oscillation frequency in the spectrograms in Fig.~\ref{fig:Time_Frequency} is largely time-independent. \\

The second dominant feature, the linearly-chirped rapid oscillations,  has its origin in a transition between the ground state and one of the higher-lying excited levels ($j=5$). The corresponding gap $\Delta_{05}$ increases linearly with $V$ (see Fig.~\ref{fig:Projections}b), its slope for $L=10$ matches well the slope of the corresponding feature in the short-time Fourier transform. The relationship between the two is $\textcolor{black}{\tfrac{\overline{b}}{v}\sim \tfrac{1}{2\pi}\tfrac{\mathrm{d}\Delta_{05}}{\mathrm{d} V}}$. The excitation to this (and others) specific state(s) occurs while driving the system through the critical point $\Delta=1$. In panel c of Fig.~\ref{fig:Projections}, we plot the projection of the time-evolved initial pure state $\psiG(t)$ (coinciding with the GS of the LL-model at $t_0=0$)  onto the $j^\mathrm{th}$-eigenstate of the time-dependent Hamiltonian $\Ham(t)$ (denoted by $\psi_t^{(j)}$) at the \textit{end of the time evolution} $t_\mathrm{max}$. Apart from the nearly-degenerate low-lying states with $j=1,2$, only the state with $j=5$ has a significant contribution, and beatings with the $\psiG(t)$ dominate the time-frequency structure of the non-equilibrium QFI. The state $\psi_t^{(j=5)}$ is dominated by spin configurations featuring two domain walls (e.g. $\ket{\su\su\sdn\su\su\sdn\su\sdn\su\sdn}$) for all values of $t$, whereas the time-evolved $\psiG(t)$, especially after crossing $\Delta=1$, becomes progressively dominated by the defect-free two N\`{e}el configurations. The projection of  $\psiG(t)$ onto the initial state at $t_0$, plotted as a dark purple line in Fig.~\ref{fig:Projections}c, exhibits an interesting dynamics consisting of a decay followed by multiple revivals. For completeness, in panel d of the same figure we show the temporal evolution of several matrix elements $b_{ij}=\left\langle \psi_t^{(i)} \right. \left| \psiG(t)\right\rangle \left\langle  \psiG(t) \right. \left|\psi_t^{(j)}\right\rangle$, i.e. the projection of $\psiG(t)$ over the instantaneous eigenstate at each $t$. These results imply that the non-equilibrium QFI in the impulsive regime is determined by the non-adiabatic excitation of multiple states at the critical point and the associated increased domain wall density.


\section{Lindblad master equation}\label{sec:Lindblad}

\begin{figure}[h]
	\centering
	\includegraphics[scale = 0.85]{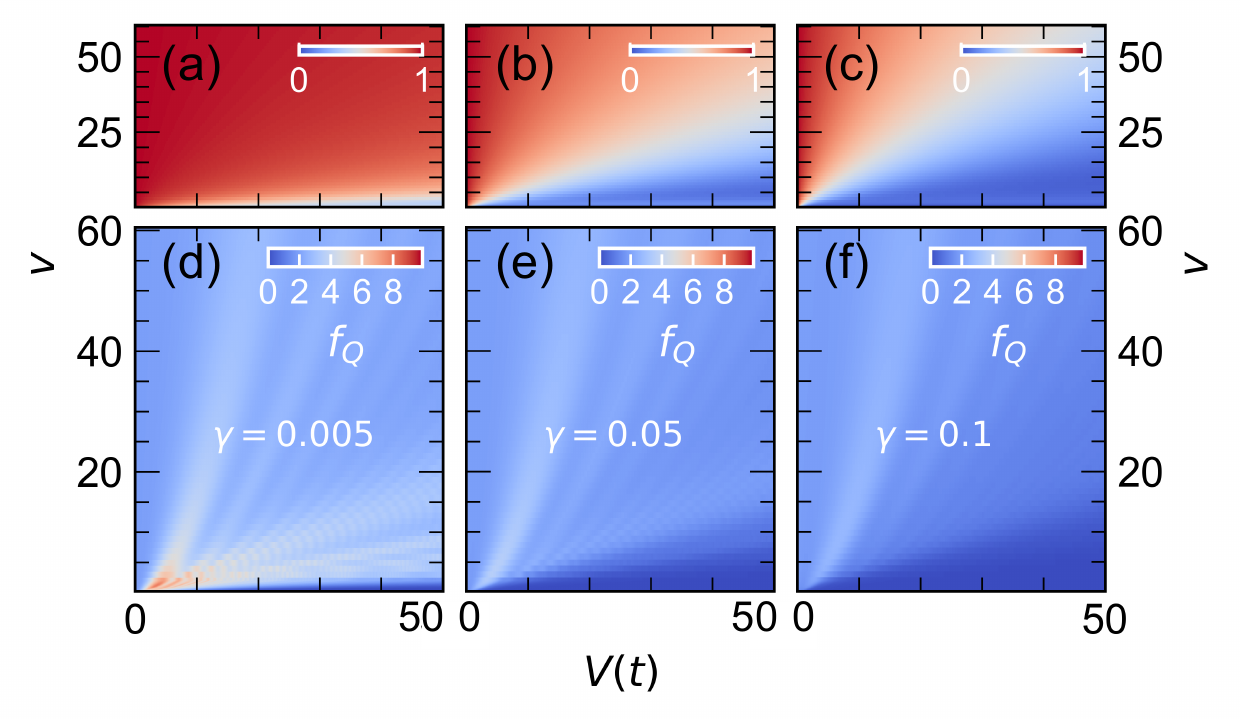}
	\caption{(a)-(c) Density matrix purity as a function of effective nearest-neighbour Coulomb repulsion and ramp speed. The decoherence rate is set to $\gamma=0.005$ in panels (a)/(d), to $\gamma=0.05$ in (b)/(e), and to $\gamma =0.1$ in (c)/(f).
	(d)-(f) Non-equilibrium QFI in the presence of decoherence for three different decoherence rates $\gamma$, displayed as a function of the effective nearest-neighbour Coulomb repulsion and ramp speed. The colormap is normalized with respect to the unitary QFI density, shown in Fig.~\textcolor{black}{2a} of the main text.   }\label{fig:Decoherence_Rates}
\end{figure}

In Fig.~\ref{fig:Decoherence_Rates}, we present calculation results showcasing the influence of the decoherence rate on the  QFI during non-unitary time evolution ($L=10$, $h_z = 0.005$). The main \textcolor{black}{non-equilibrium QFI features, i.e. the broad QFI maximum and the subsequent oscillatory dynamics, are preserved for coupling strengths up to $\gamma\sim 0.1$, and progressively fade when $\gamma> 0.1$}. The dissipative Lindblad jump operator acting on each site $l$ is given by $\Lb^{(l)}\equiv \Lb_{z} = \hat{\sigma}^z_l$. \\
\par The QFI in the presence of decoherence is calculated by first propagating the density matrix $\rhoDM(t)$ obeying the initial condition:
\begin{equation}
    \rhoDM(t_0) = \ket{\psiG(t_0)}\bra{\psiG(t_0)}. \label{eq:rhoDM_t0}
\end{equation}
The QFI is obtained as~\cite{Hauke2016measuring}:
\begin{equation}
    \FQ(t) = 2 \sum_{i,j}\frac{\left(\epsilon^i_t-\epsilon^j_t\right)^2}{\epsilon^i_t+\epsilon^j_t} 
    \left|\bra{\varphi^i_t}\Oop_\pi\ket{\varphi^j_t}\right|^2\label{eq:QFI_DM},
\end{equation}
where $\epsilon^i_t$ and $\ket{\varphi^i_t}$ are the eigenvalues and the eigenvectors of $\rhoDM$ at time $t$.

\section{Kibble-Zurek analysis of the nonequilibrium QFI}\label{sec:Limits}
\begin{figure}[h]
	\centering
	\includegraphics[scale = 0.85]{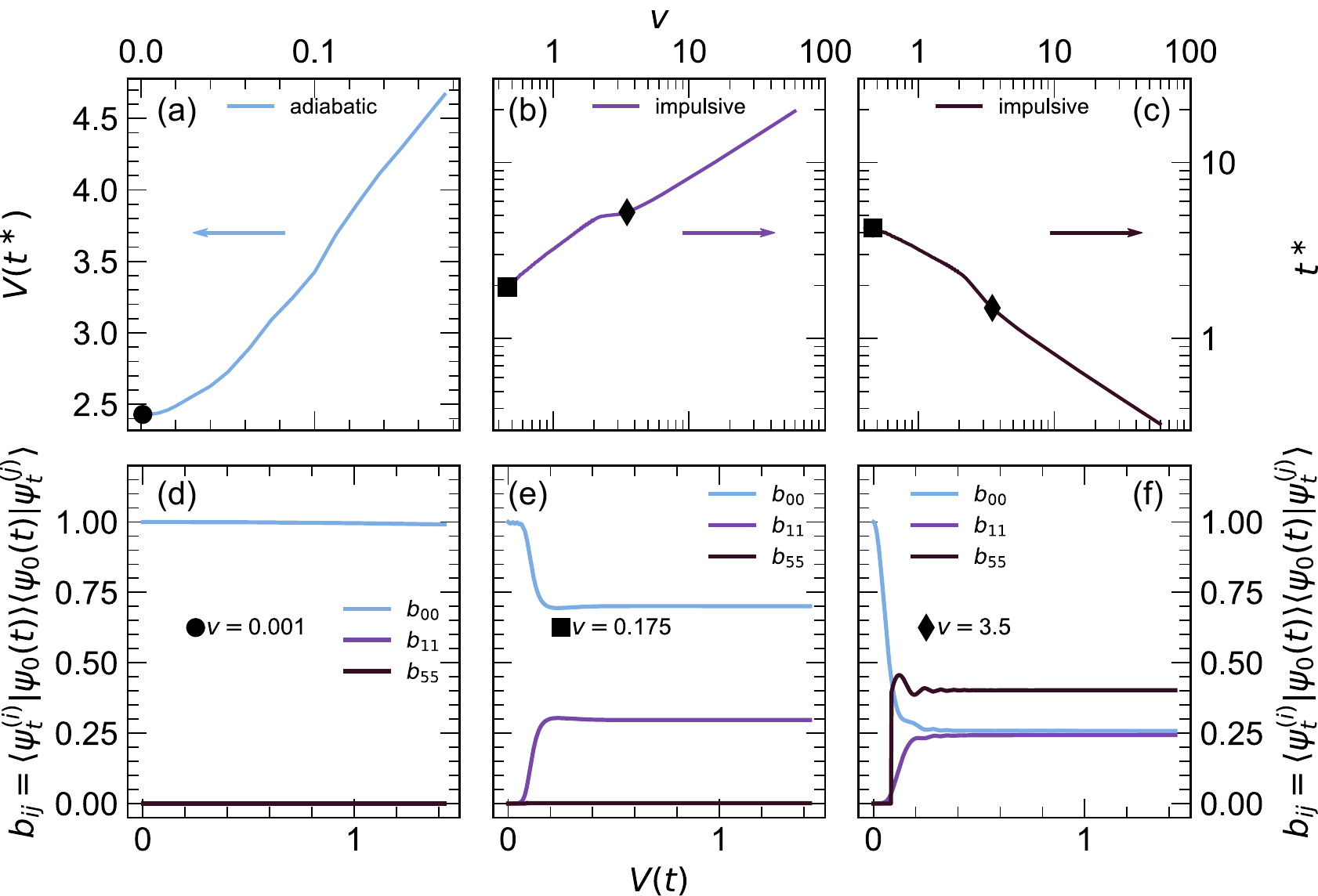}
	\caption{(a)-(c) Positions of the maxima of the time-dependent QFI density $f_Q$ as a function of the ramp velocity, displayed as a function of the effective nearest-neighbour Coulomb repulsion [panels (a) and (b)] or time (c). Points falling into the adiabatic and (partially) to the ``intermediate'' regime are shown in panel (a). Panels (b) and (c) show the ``intermediate'' and the non-adiabatic regions on a double-logarithmic scale. The onset of the non-adiabatic regime is manifested by a marked change in the slope of the curves (see text). The symbols indicate the positions of the three different ramp velocities corresponding to the data in the bottom row. (d)-(f) Time-dependent matrix elements $b_{ij}$, given by products of overlaps of the time-propagated $\psiG(t)$ with the eigenstates of the \textit{instantaneous} Hamiltonian  $\Ham(t)$ at time $t$, shown for three different velocities (see legends). }\label{fig:Limits}
\end{figure}

The nonequilibrium QFI dynamics (see. Fig.~\textcolor{black}{2a} of the main text) of the spinless fermion chain subject to a linear ramp can be broadly divided into three regions depending on the ramp velocity. These regimes can be delineated by tracking the time ($\txt$) or the effective  nearest-neighbour Coulomb interaction ($\Jxt$) which maximize the  time-dependent $f_Q$ at a given ramp speed $\mathit{v}$. The resulting curve (s. white line in Fig.~\textcolor{black}{2a} or Fig.~\ref{fig:Limits}a-c) effectively tracks the position of the main QFI ``crest'' as it propagates towards larger $\Jxt$ (smaller $\txt$) on increasing the ramp speed. The positions of the maxima are extracted by fitting a skewed Voigt distribution to the time profile of $f_Q$ at each velocity in a region centered around the main peak.   \\

\par The two limiting cases are defined by the adiabatic (\textcolor{black}{$v<0.03$}, cf.\ Fig.~\ref{fig:Limits}a) and the nonadiabatic ($v>3.5$, cf.\ Fig.~\ref{fig:Limits}b-c) limits, separated by an intermediate region. Whereas the onset of the non-adiabatic region is rather sudden and well-marked by a ``kink'' in the $\mathit{v}-\Jxt$ resp. $\mathit{v}-\txt$ plots in Fig.~\ref{fig:Limits}b-c, the separation between adiabatic and intermediate regions is rather fluid.  In the deep adiabatic regime, the position of the $f_Q$-maximum is weakly dependent on the driving speed, and the driven system essentially traces out the equilibrium phase diagram. This regime (illustrated for $v=0.001$ in Fig.~\ref{fig:Limits}d) holds as long as excitations to the first excited state on crossing the critical point $\Delta=1$ can be neglected, i.e. $b_{ij}\approx \delta_{0,i}\delta_{0,j}$ for the entire time evolution. Examining the time-resolved data underlying Fig.~\textcolor{black}{2a}, we estimate a value of $\vcrf\sim0.03$ for the maximum ramp speed that warrants purely adiabatic evolution before oscillations due to excitations to the first excited state develop.
Invoking simple Landau-Zener (LZ) arguments, we can estimate the corresponding excitation probability to the first excited state as $P_{01}=1-e^{-\frac{\pi\tilde{\delta}\Delta_{01}^2}{2v}}$, where $\tilde{\delta}$ is a constant related to the rate of change of the gap at $\Delta=1$ as a function of the critical parameter $V(t)$. The threshold for the breakdown of adiabaticity in the LZ picture ($\Delta_{01}^2\approx v|\mathrm{d}\Delta_{01}/\mathrm{d}V|$), however,  largely overestimates $\vcrf$, indicating a more complex behaviour.  \\

\par For ramp velocities above $\vcrf$ but not exceeding $\vcrs$, the dynamics falls into the intermediate regime, characterized by an extensive scaling of the QFI density $f_Q\sim L$ in the unitary evolution case. Due to the rapidly oscillating and saturating behaviour of $f_Q$, the corresponding maxima could not be extracted for velocities ranging from $0.175$ up to shortly before the onset of the non-adiabatic region $\vcrs\sim 3.5$ (cf. square/diamond symbols in Fig.~\ref{fig:Limits}b-c). Here, the dynamics are mainly governed by the ground and the first excited states of the system, as one can deduce from the dominant $b_{ij}$-coefficients in Fig.~\ref{fig:Limits}e. Recalling that in the presence of the staggered magnetic field of amplitude $h_z$ the ground state at $t_0$ is dominated by one of the two N\'e{e}l configurations, creating excitations across the gap mixes the  wavefunction with the state where the other  N\'e{e}l is dominant. The resulting superposition contains these two components with a similar amplitude, hence the growth of the entanglement. \\

\begin{figure}[h]
	\centering
	\includegraphics[scale = 0.85]{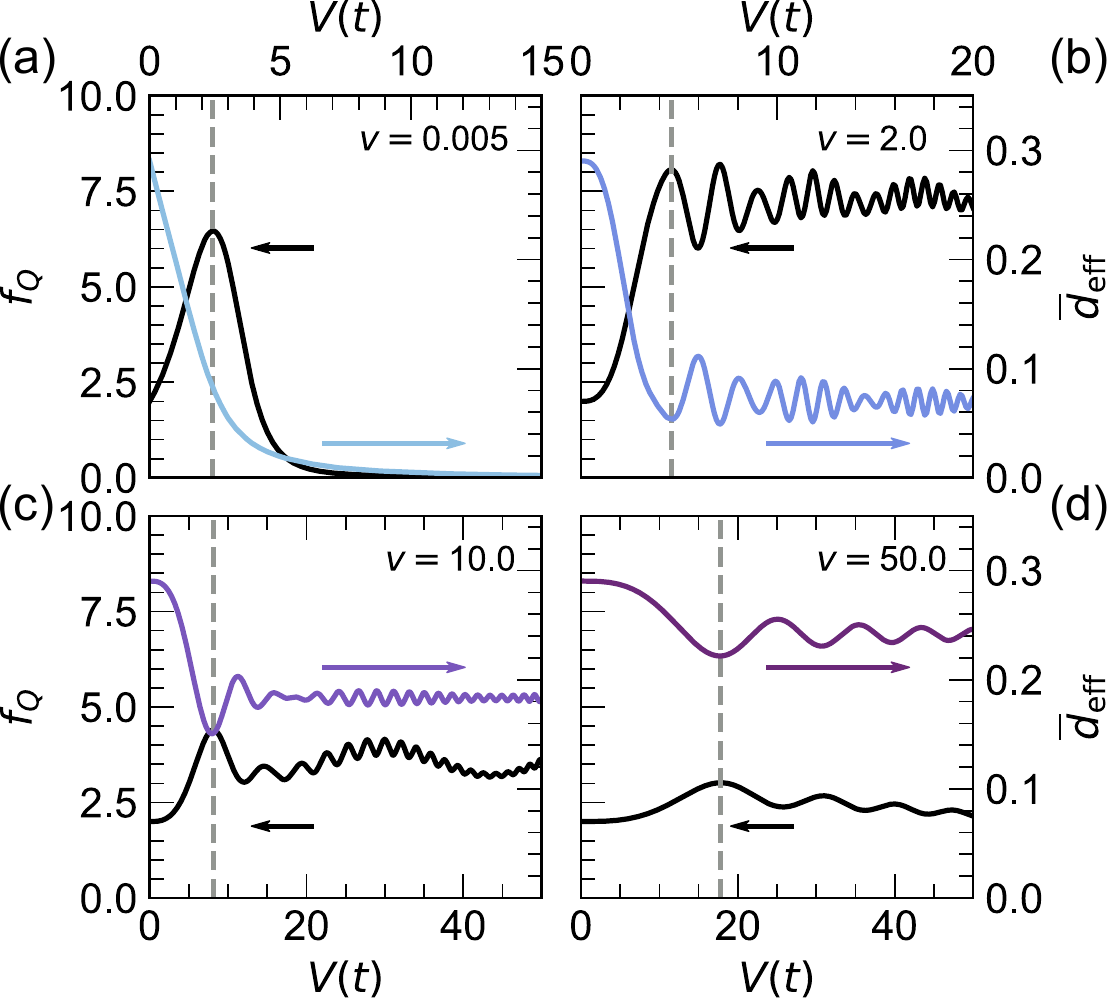}
	\caption{\textcolor{black}{Relation of the dynamical QFI density to the time evolution of the integrated density of states for four representative ramp velocities: $v=0.005$ (a), $v=2.0$ (b), $v=10.0$ (c), and $v=50.0$ (d). Each panel shows the integrated density of defects (color) plotted against the corresponding $f_Q (t)$ at that velocity (black). }
}\label{fig:Defects_Extrema}
\end{figure}

\par The non-adiabatic regime has a sharply delineated onset, marked by a rapid change of the slopes of $\Delta v / \Delta \Jxt$ resp.\ $\Delta v / \Delta \txt$. The nonadiabatic limit commences when the probability of an excitation to the higher-lying states (in particular $j=5$ for the $L=10$ chain) exceeds the corresponding excitation fraction to the first excited state $j=1$ (cf. Fig.~\ref{fig:Limits}f), leading to proliferation of defects. \textcolor{black}{As visible in Fig.~\ref{fig:Defects_Extrema}, the dynamics of the defect proliferation closely follows the dynamical QFI, and we therefore we track the temporal maximum of $f_Q$ as a function of the ramp velocity and interpret the results with the aid of a KZ scaling analysis.}  In this regime, the maximum of the QFI density follows a regular behaviour, and its functional dependence can be extracted by fitting the corresponding slopes in Fig.~\ref{fig:Limits}b-c: $\Jxt\propto \sqrt{v}$ resp.\ $\txt\propto v^{-1/2}$. The exact values of the slopes are \textcolor{black}{$0.5295\pm0.0009$} and \textcolor{black}{$-0.5937\pm0.0019$}, respectively. With the aid of the critical exponents of the system at $\Delta=1$, we can relate the scaling of the entanglement jet to the corresponding scaling of the density of defects in a Kibble-Zurek type of analysis: 
    \begin{equation}
        d_\mathrm{KZ}\propto \frac{a_0}{\xi_0}\left(\tau_0 v\right)^{1/2}\label{eq:dKZ},
    \end{equation}
where $a_0$, $\xi_0$, $\tau_0$ are the characteristic length scale, correlation length, and relaxation rate of the system. Note that a KZ-analysis has a limited validity at the phase transition at the $\Delta=1$ critical point is of BKT type instead of a second-order one. 

\section{Nonequilibrium QFI in presence of next-nearest-neighbor interactions}\label{sec:Next_nearest_neighbor}
\begin{figure}[h]
	\centering
	\includegraphics[scale = 0.85]{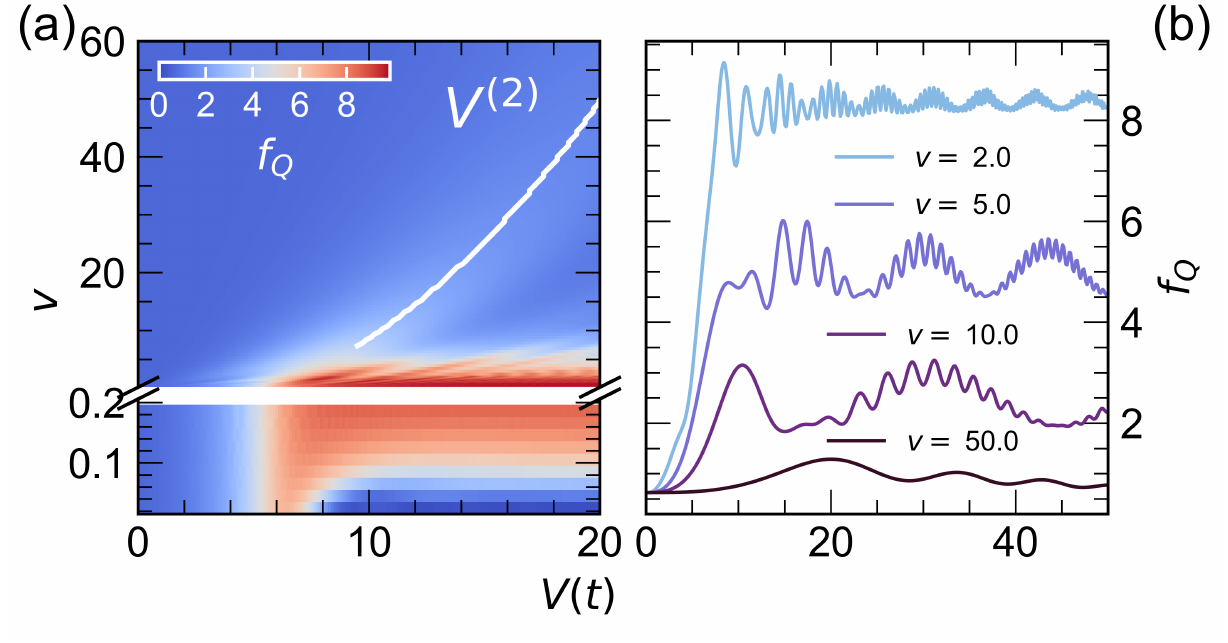}
	\caption{(a) Non-equilibrium QFI for a Heisenberg chain featuring additional next-nearest neighbour interaction term $\Ham_{(2)}$ as a function of the nearest-neighbour Coulomb repulsion and the ramp speed. (b) Selected 1D line cuts for several ramp velocities.}\label{fig:Second_Neighbours}
\end{figure}

Finally, we corroborate the stability (or universality) of the nonequilibrium QFI features by also considering the role of additional interactions, that break integrability in the infinite system~\cite{patrick2019}. We specifically study the dynamical QFI for the spinless fermion chain at half-filling (Eq.~\eqref{eq:Hamiltonian_fermionic}) with the inclusion of next-nearest-neighbor interactions of the form $\Ham_{(2)} = V^{(2)}\sum_l \tilde{n}_l\tilde{n}_{l+2}$. The results presented in Fig.~\ref{fig:Second_Neighbours}c for a second-order coupling of $V^{(2)}=2$ illustrate that the main traits of the nonequilibrium QFI dynamics are preserved. Importantly, this observation underscores the generic character and robustness of our main findings and suggests that they should be observable across many different experimental platforms, irrespective of microscopic details.

\section{Dynamical QFI and order parameter}\label{sec:Mstagg}

\begin{figure}[h]
	\centering
	\includegraphics[scale = 0.85]{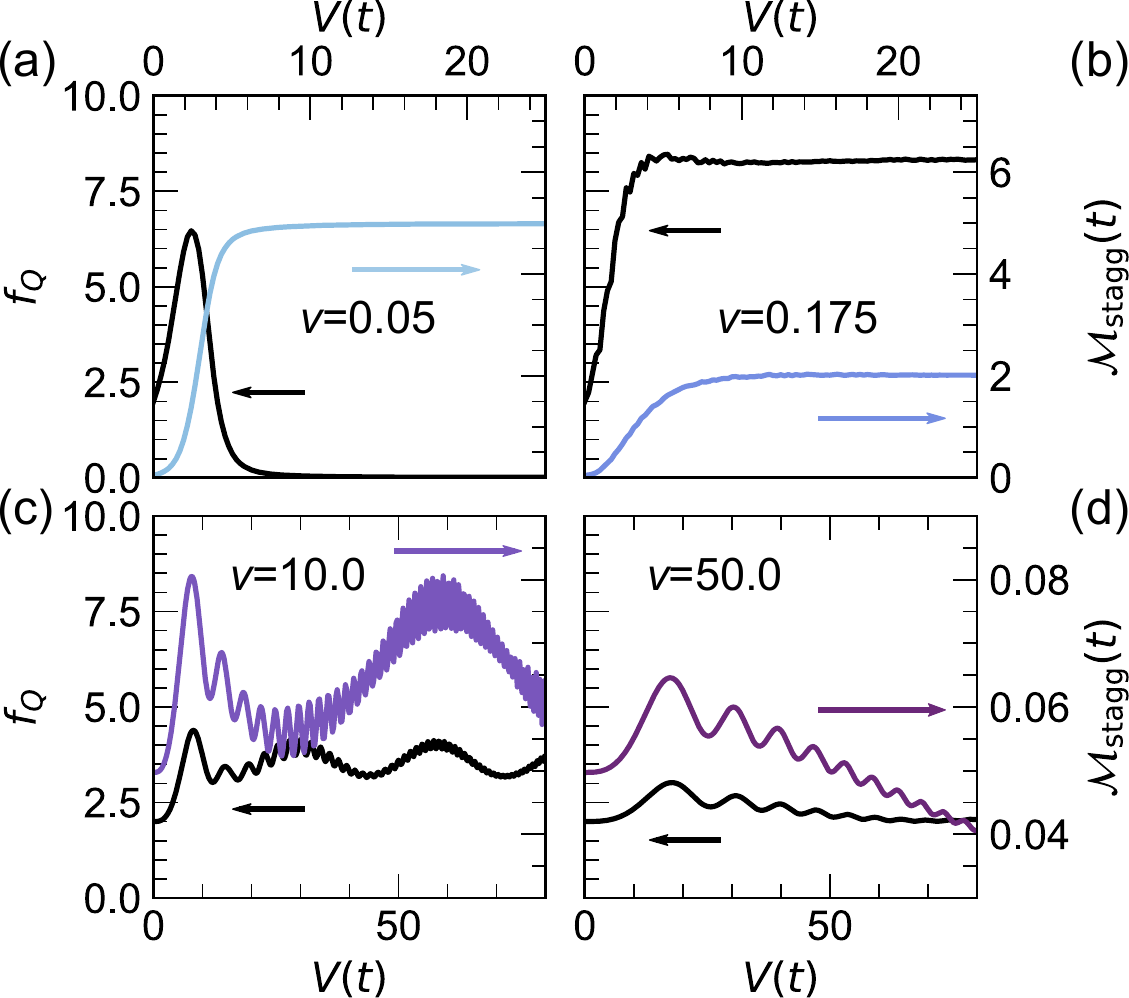}
	\caption{\textcolor{black}{Relation of the dynamical QFI density to the time evolution of the order parameter of the LL $\rightarrow$ CDW phase transition for four representative ramp velocities: $v=0.05$ (a), $v=0.175$ (b), $v=10.0$ (c), and $v=50.0$ (d). Each panel shows the staggered magnetization $\cMstagg (t)$ (color) plotted against the corresponding $f_Q(t)$ at that velocity (black). The calculations are performed for a chain of size $L=10$ and a staggered magnetic field of $h_z=0.005$. }}\label{fig:Staggered_Mag}
\end{figure}

\textcolor{black}{In this section, we compare the dynamical QFI density to the time evolution of the order parameter of the LL $\rightarrow$ CDW phase transition, i.e.\ the staggered magnetization $\cMstagg (t)$. In Fig.~\ref{fig:Staggered_Mag}, we present the dynamical behaviour of $\cMstagg(t)$ for a $L=10$ chain subject to a time-dependent nearest-neighbor linear interaction ramp $V(t)$ and including a small staggered magnetic field $h_z=0.005$. We choose four representative ramp velocities spanning the adiabatic, intermediate, and impulsive interaction regimes. 
The behaviour of the two quantities $f_Q (t)$ and $\cMstagg (t)$ displays qualitative similarities in the intermediate and impulsive regions, but vast discrepancies in the adiabatic limit.}\\

\textcolor{black}{Due to the presence of the additional staggered magnetization term $h_z$, $\cMstagg (t)$ saturates to a finite asymptotic value in the CDW phase and does not exhibit a maximum at the critical point for adiabatic ramps (cp.~Fig.~\ref{fig:Staggered_Mag}a). In the absence of an additional staggered magnetic field, $\cMstagg (t)$ evaluates to zero irrespective of the ramp velocity due to the canceling contributions from the degenerate N\'{e}el states.}\\

\textcolor{black}{Overall, the staggered magnetization reaches its maximum value for adiabatic ramps. In the intermediate, ``critical'' region, where $f_Q$ is superextensive ($f_Q\sim L$), $\cMstagg$ saturates at a lower value with respect to the adiabatic case (Fig.~\ref{fig:Staggered_Mag}b). In the impulsive limit (panels c and d), the position of the first QFI crest is mirrored in a maximum in the staggered magnetization. The subsequent fast oscillations of the two quantities are phase-synchronized at first, but slowly go out of phase with progressing time. Furthermore, the periodicity of the slow oscillatory motion is completely different in both cases. More importantly, for fast ramps $v\ge 40$, the staggered magnetization remains very small ($\cMstagg < 0.05$) throughout the entire time evolution. In contrast, the QFI density reaches values exceeding the classical  bound for all considered ramp velocities. }

\section{QFI Dynamics at finite temperature}\label{sec:finiteT}
\begin{figure}[h]
	\centering
	\includegraphics[scale = 0.85]{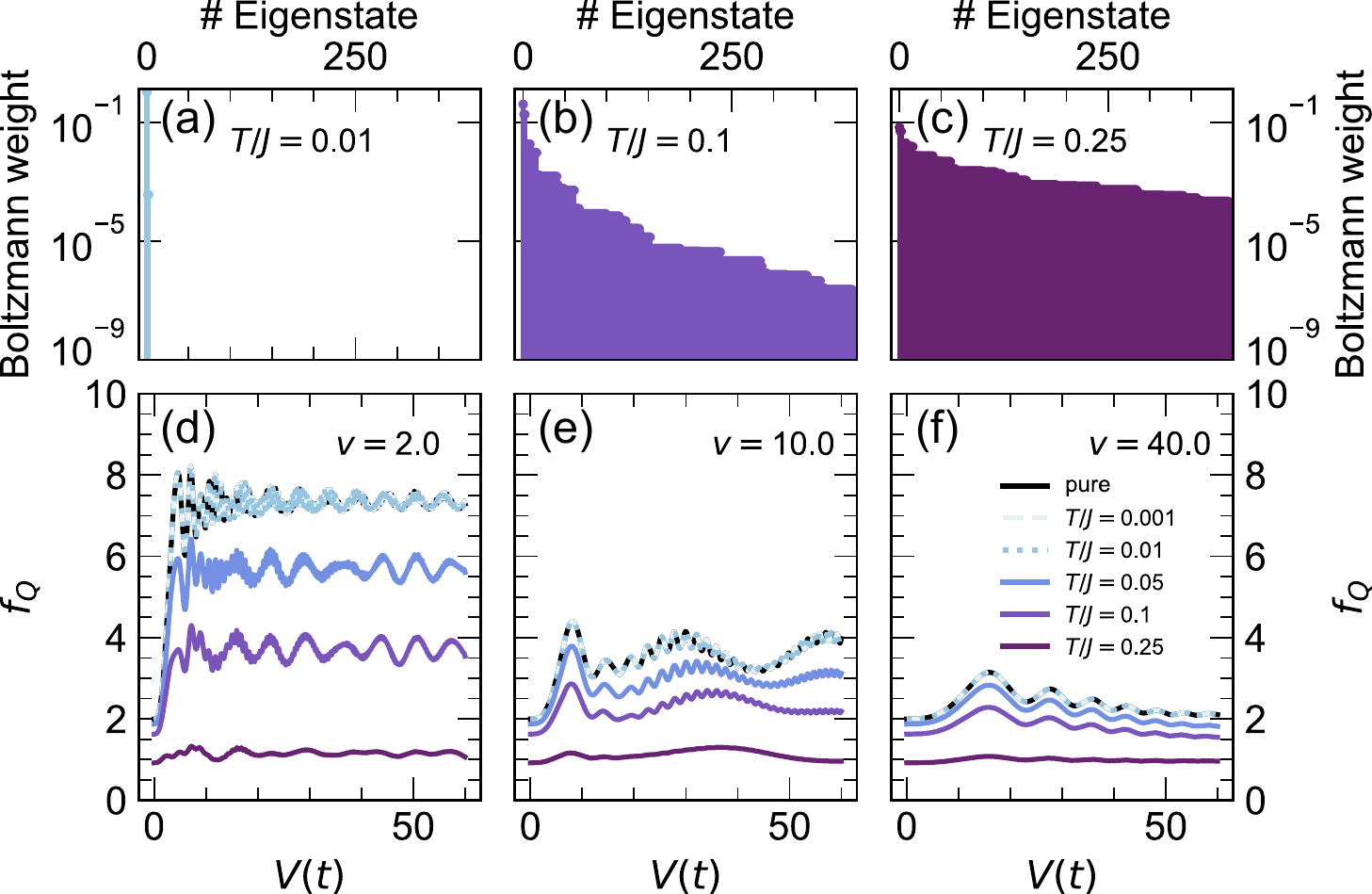}
	\caption{\textcolor{black}{Non-equilibrium QFI for thermal states at various temperatures. Panels (a), (b), and (c) display the distribution of the Boltzmann factors over the eigenstates of a $L=10$ spinless fermionic chain for three different temperatures $T/J$: $0.01$, $0.1$, and $0.25$. Panels (d) - (f) show the non-equilibrium QFI density $f_Q$ for selected temperatures for the following three velocities: $v=2.0$ (d), $v=10.0$ (e), and $v=40.0$ (f). Note that the dynamics at $T/J = 0.001$ and $T/J=0.01$ follow closely the pure state dynamics such that the three curves are nearly coincident.}}\label{fig:finiteT}
\end{figure}

\textcolor{black}{While the numerical results included in the main follow the evolution of a pure initial state, this assumption is rarely fulfilled in macroscopic physical systems, where the starting point is often represented by a mixed state at thermal equilibrium. To demonstrate the validity of our results also in the experimentally-relevant situation of mixed initial states, we have calculated the dynamical evolution of the QFI under a linear interaction ramp starting from a thermal ensemble. In Fig.~\ref{fig:finiteT}, we present numerical results for the Liouville-von-Neumann evolution of a $L=10$ spinless fermionic chain subject to a nearest-neighbour Coulomb interaction quench at varying ensemble temperatures (reported in terms of the ratio $T/J$) and for three representative ramp velocities. From these results, we can deduce the following: first and most important, the main features of the pure-state QFI dynamics, i.e. the initial rise of the system entanglement, the broad maximum, and the subsequent oscillatory structure – are preserved in this mixed-state case, and the initial population of higher-lying states generally reduces the QFI density.
For temperatures up to $T/J\sim0.01$, this decrease is negligible. At higher temperatures (up to $T/J\sim0.1$), the QFI is increasingly suppressed (while still exceeding the classical bound) and the features of its temporal evolution (such as the phase of the oscillations) are preserved. Above $T/J\sim0.2$ the QFI decreases below the classical bound, thus preventing its use as a multipartite entanglement witness.}\\

\textcolor{black}{These thermal states are directly relevant to real condensed-matter systems, where the mixed-state QFI can be extracted via an integral of the dynamical susceptibility \cite{Hauke2016measuring} as measured e.g. in inelastic neutron scattering experiments \cite{Scheie2021witnessing,Laurell2021quantifying}. Typical values of the exchange constant $J$ for the Mott-insulating model systems considered in this work are on the order of few tenths of eV: e.\ g.,\  $0.1$~eV for $\mathrm{ET-F_2TCNQ}$~\cite{Wall2011quantum}, $0.17$~eV for $\mathrm{Sr_2CuO_3}$~\cite{Iwai2003ultrafast}, or $0.26$~eV for $\mathrm{[Ni\left(chxn\right)_2Br]Br_2}$~\cite{Iwai2003ultrafast}. Given the thermal energy of $0.025$~eV, this leads to $T/J$-values of $0.1 - 0.25$ at room temperature. At cryogenic temperatures ($< 30$~K), $T/J$ can be suppressed by an order of magnitude. Thus, depending on the exact material, the mixed state QFI for a thermal initial state under experimentally-relevant temperature conditions closely follows the prototypical pure state evolution discussed in the main text.}

%